\documentstyle[epsf]{SolarPhysics}
\def\gapprox{\;\rlap{\lower 2.5pt            
 \hbox{$\sim$}}\raise 1.5pt\hbox{$>$}\;}       
\def\lapprox{\;\rlap{\lower 2.5pt            
 \hbox{$\sim$}}\raise 1.5pt\hbox{$<$}\;} 
\def\N{\,{\rm I\kern-.20em N}}
\begin{document}

\begin{article}
\begin{opening}
\title{High Spectral Resolution Observation of Decimetric Radio Spikes Emitted by Solar Flares\ \ --\ \ First Results of the Phoenix-3 Spectrometer}
\author{Arnold O. \surname{Benz}$^1$}
\author{Christian \surname{Monstein}$^1$}
\author{Michael \surname{Beverland}$^1$}
\author{Hansueli \surname{Meyer}$^1$}
\author{Bruno \surname{Stuber}$^2$}
\institute{$^1$ Institute of Astronomy, ETH Zurich, 8093 Z\"urich, Switzerland (e-mail: benz at phys.ethz.ch)\\
$^2$ Institute of Automation, University of Applied Sciences FHNW, 5210 Windisch, Switzerland\\}
\runningtitle{Decimeter Radio Spikes}
\runningauthor{Arnold O. Benz et al.}
\date{Received: xxx; accepted: xxx}
\begin{abstract} 
A new multichannel spectrometer, Phoenix-3, is in operation having capabilities to observe solar flare radio emissions in the 0.1 - 5 GHz range at an unprecedented spectral resolution of 61.0 kHz with high sensitivity. The present setup for routine observations allows measuring circular polarization, but requires a data compression to 4096 frequency channels in the 1 - 5 GHz range and to a temporal resolution of 200 ms. First results are presented by means of a well observed event that included narrowband spikes at 350 - 850 MHz. Spike bandwidths are found to have a power-law distribution, dropping off below a value of 2 MHz for full width at half maximum (FWHM). The narrowest spikes have a FWHM bandwidth less than 0.3 MHz or 0.04\% of the central frequency. The smallest half-power increase occurs within 0.104 MHz at 443.5 MHz, which is close to the predicted natural width of maser emission. The spectrum of spikes is found to be asymmetric, having an enhanced low-frequency tail. The distribution of the total spike flux is approximately an exponential. 

\end{abstract} 
\end{opening}
\section{Introduction}
The observation of coherent solar radio emission during flares requires high resolution in frequency. Here we present first results of a new spectrometer that extends the limits of frequency resolution and sensitivity by an order of magnitude. We show an example of the recording of narrowband spikes.

Spikes in decimeter radio waves have the shortest bandwidth requiring the highest frequency resolution of the coherent flare radio emissions. They have first been noted for their extremely short duration (Dr\"oge, 1967, 1977) and association with gyrosynchrotron radiation (Slottje, 1978) and hard X-ray flare emission (Benz, 1985). Spikes appear in large groups of up to several thousand single peaks. The duration of a single spike is on the average 73$\pm$22 ms at 362 MHz and decreases about linearly with inverse frequency (G\"udel and Benz, 1990; Rozhansky, Fleishman, and Huang, 2008; Sirenko and Fleishman, 2009). A different class of spikes at 200 - 450 MHz was reported by Magdalenic et al. (2006) with durations as small as 4 ms. Spike durations as short as 2.6 ms at 1.4 GHz, 3.0 ms at 2.695 GHz, and 2.5 ms at 5 GHz have been reported by Dabrowski et al. (2005), P. Zlobec (private communication, 2009), and Rozhansky, Fleishman, and Huang (2008), respectively. 

Later it became clear that spikes also have extremely narrow bandwidth (review by Benz, 1986), which became an important classification characteristic of spikes, contrasting them from short but broadband pulsations also observed in decimeter waves. The bandwidth of spikes increases with frequency. Csillaghy and Benz (1993) derived an average relation between the FWHM bandwidth $\Delta \nu_{\scriptscriptstyle\rm FWHM}$ and center frequency $\nu$ in the frequency range from 300 to 8500 MHz,
\begin{equation}
\Delta \nu_{\scriptscriptstyle\rm FWHM} = 0.66\ \nu^{0.42}\ \ \ .
\end{equation}
(All frequencies in MHz). Contrary to spike duration, the spike bandwidths show large variations at a given frequency. The smallest bandwidths in terms of the ratios of $\nu / \nu_{\scriptscriptstyle\rm FWHM}$ for individual spikes at 300 MHz and 900 MHz of 0.41\% and 0.17\%, respectively, were reported by Messmer and Benz (2000). Their observations had a resolution of 1 MHz, the best to date at decimeter waves below 1 GHz. Above 1 GHz, the minimum bandwidth was found limited by instrumental resolution.  Upper limits for the minimum $\Delta \nu_{\scriptscriptstyle\rm FWHM} / \nu$ ratios of 0.23\% (Csillaghy and Benz, 1993), and 0.2\% (Rozhansky, Fleishman, and Huang, 2008) were reported. At 3.47 GHz, St\"ahli and Magun (1986) reported relative bandwidths between less than 0.014\% and more than 5.8\%. Apparently, spikes have no characteristic bandwidth.  This is corroborated by the power law that results from Fourier transforming spectral scans through spike events (Karlicky, Sobota, and Jiricka, 1996; Karlicky, Jiricka, and Sobota, 2000).

The bandwidth is an important diagnostics for the emission process. There are good reasons to assume that the source is small (see below), thus the brightness temperature is extremely high, suggesting a coherent emission process. Emissions by beam instability and electron cyclotron maser (ECM) instability are prime candidates. Both are able to produce a $\Delta \nu_{\scriptscriptstyle\rm FWHM} / \nu$ ratio of a few 0.01\% (Csillaghy and Benz, 1993; Benz, 2002; Fleishman, 2004a). 

Broadening of the EMC emission due to wave turbulence has been studied by Fleishman (2004b). He finds that the broadening is proportional to the ratio of the mean square wave magnetic field to the square of the background magnetic field. This is consistent with the reported minimum values of $\Delta \nu_{\scriptscriptstyle\rm FWHM} / \nu$. If the background magnetic field $B$ changes systematically over the dimension of the source, the broadening by the non-uniformity of the magnetic field dominates and the bandwidth of EMC radiation is
\begin{equation}
{\Delta \nu_{\scriptscriptstyle\rm FWHM} \over \nu} = {\Delta B\over B} \approx {\Delta L\over H_B}\ \ \ ,
\end{equation}
where $\Delta L$ is the size of the radio source and $H_B$ is the magnetic scale length. 

The distribution of spike peak fluxes vs. number is related to the cause and circumstances driving the involved waves. 
\begin{itemize}
\item If the spike emission is caused by an exponentially growing instability but stopped in growing after some random duration, the peak flux distribution is expected to be a {\it power law}. The same distribution results from a system in self-organized criticality (Lu and Hamilton, 1991; Georgoulis and Vlahos, 1998; Charbonneau et al., 2001; Nishizuka et al., 2009). 
\item Stochastic growth in a large number of uncorrelated regions of maser instability yield {\it log-normal} distributions as found e.g. in interplanetary type III bursts (Roelof and Pick, 1989; Cairns and Robinson, 1998), in the radio emission from the bow shock of the Earth (Cairns and Robinson, 1999), and from the termination shock (Kuncic, Cairns, and Knock, 2004). 
\item If the velocity distribution is  unstable but free energy continuously replenished, spiky emission results. The flux of individual elementary bursts is distributed {\it exponentially} (Robinson, Smith, and Winglee, 1996). 
\end{itemize} 
Reports on observed distributions are contradictory. Exponential distributions of spike peak fluxes have been reported by Robinson, Smith, and Winglee (1996), Aschwanden, Dennis, and Benz (1998), Meszarosova et al. (2000), Isliker and Benz (2001), and Rozhansky, Fleishman, and Huang (2008). On the other hand, Nita, Fleishman, and Gary (2008) found a power-law distribution after making the spike detection algorithm more effective for small bursts. Furthermore, Meszarosova et al. (2000) report power-law distributions in events with sparse spike occurrences. Cairns and Benz (unpublished) found a power-law distribution in metric spikes. Isliker and Benz (2001) demonstrate mathematically that unresolved measurements in time or spectrum may turn a power law into an apparently exponential distribution.  

Here we present first results of a new radio spectrometer, Phoenix-3, dedicated to solar flare observations. The instrument and the current routine observing mode are described in Section 2. Data taken during a test observation are presented in Section 3. They include a serendipitously observed event of decimetric spikes which is analyzed, presented in detail and discussed in Section 4. The limb event of GOES class M3 is the most interesting event observed to date. These first Phoenix-3 data, presented here, are of general interest for the understanding of narrowband spikes. The high spectral resolution allows to address the above questions on minimum bandwidth and peak flux distribution. Conclusions on both the new instrument and its first observations are given in Section 5.

 \section{The Phoenix-3 Spectrometer}
 
Phoenix-3 is the latest addition to the series of solar radio spectrometers at ETH Zurich that started in the early 1970s (Perrenoud, 1982). The original concept of digital spectrometers was improved and further developed in Phoenix (Benz et al., 1991) and Phoenix-2 (Messmer, Benz, and Monstein, 1999). The observatory in Bleien 50 km west of Zurich, the 7m parabola, log-periodic antenna, and parts of the software were taken over from one spectrometer generation to the next. During the past three decades many technological improvements have been made. 
 
\begin{figure}
\begin{center}
\leavevmode
\mbox{\hspace{0.0cm}\epsfxsize=12cm
\epsffile{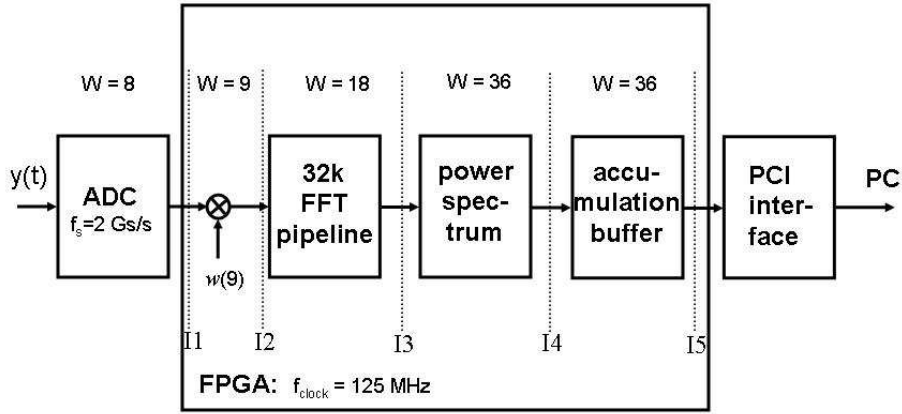}}
\end{center}
\caption[]{Schematic drawing of the Phoenix-3 spectrometer unit. ({\it i}) AC240 hardware, consisting of the sampler (ACD) plus Field-Programmable Gate Array (FPGA). ({\it ii}) Software: spectral analysis by Fast Fourier Transformation (FFT), power spectrum formation and integration.}  
  \label{fig:1}
\end{figure}

Contrary to its frequency-agile predecessors, Phoenix-3 is a multi-channel spectrometer observing in all channels all the time. The core of the spectrometer is a digital sampler combined with a fast processor, an FPGA chip (XC2VP70), that fast-Fourier transforms the signal into a spectrum. The hardware is a commercially available device, AC240, produced by Acqiris SA, Geneva (now Agilent Technologies). The FFT was developed and tested by our research institutes. Figure 1 displays a schematic overview. The initial digitization (ADC) is 8 bit wide. This word length $W$ is increased in the course of the analysis pipeline from one step to the next (indicated by vertical dotted line in Figure 1) to reduce the effects of truncation. Sampling the analogue input signal $y(t)$ at a rate of 2 GHz, the 16 parallel streams are organized with a clock rate of 125 MHz at interface I1 within the FPGA. Next, a window multiplication, $w[9]$, is applied with an output word length of 9 bit. The word length is then extended to 18 bit within the 32$\:$768 point pipeline. At interface I3, 16k complex spectral values $X(\nu)$ are squared and transformed into real numbers yielding the power spectrum (I4). The spectral values are then integrated, storing the results in a buffer of 36 bit word length. The number of accumulation cycles is programmable. Finally at I5, the AC240 device outputs 16$\:$384 frequency channels each having a bandwidth of 61.0 kHz, providing a bandwidth of 1 GHz. 

The implementation and qualification of the FFT spectrometer has been described in Benz et al. (2005). In the mean time the FFT has been improved to minimize spurious steps in the background signal and other effects of digital signal processing. Most of the improvements were achieved by optimizing the increase of the word length in the course of the pipeline (see $W$ parameter in Figure 1). The maximum dynamic range is 48 dB, as given by the 8 bit sampling. The spurious-free dynamic range was measured 37 dB, and the half-power reception bandwidth of a narrowband signal was found to be 54 kHz. To reduce aliasing, various filter functions $w[9]$ can be inserted in addition to a rectangular filter (no filtering). We usually use the Blackman filter. It increases the effective bandwidth to 100 kHz.  

The Phoenix-3 spectrometer is an application of the AC240 device and its FFT software for solar radio observations. The range of observing frequencies covers 1 to 5 GHz. It is stepped through in four 1 GHz steps. The receiver operates according to the double heterodyne principle. Four local oscillators shift the incoming radio signal to a fixed band from 8 - 9 GHz, where it is filtered and from where it is mixed down into the 0 - 1 GHz baseband to be sampled by the AC240 and fed into the FFT pipeline. The two circular polarizations are retrieved from the linear feed via a 90$^\circ$ hybrid. They follow separate paths, and are sampled independently by the AC240 device. The data are recorded linearly in flux density and stored locally. The Allan time of the AC240 device was measured 2000 s in the lab, and 100 s in field observations including the preamplifier and the receiver. The system noise temperature is of the order of 300 K throughout most of the band. It is dominated by the broadband preamplifier.

In routine observations, the Phoenix-3 spectrometer currently observes 65$\:$536 frequency channels per 200 ms. The integration time per channel and polarization mode is 25 ms. The data rate of 327 kByte/s is too large to be transmitted to the institute. The data are compressed on-line in the following way: First, 25\% of the channels having the highest average flux density are deleted assuming that they are interfered by terrestrial transmitters. Also, the lowest 25\% are deleted corresponding to channels at the edge of the band. Then 8 adjacent channels are integrated, leaving 1024 channels per GHz. A more sophisticated scheme to reject interference is used in the data analysis of Section 3. An on-line implementation is in development.

Calibration is made with the quiet sun using the parameterized flux density $F_{\odot q}$
\begin{equation}
F_{\odot q}\   =\ 1.94\cdot 10^{-4} \nu^{1.992}\ ,\ \ \ {\rm 30\ -\ 350\ MHz},\\
\label{fluxfunc}
\end{equation}
\begin{equation}
F_{\odot q}\   =\ 8.45\cdot 10^{-1} \nu^{0.5617}\ ,\ \ {\rm 350\ -\ 6000\ MHz},\\
\label{fluxfunc}
\end{equation}
(Benz, 2009), where $\nu$ is the frequency in MHz and $F$ is in solar flux units $[$1 sfu = $10^{-19}$ erg s$^{-1}$ cm$^{-2}$ Hz$^{-1} = 10^{-22}$ W Hz$^{-1}$ m$^{-2}]$. The slowly varying component of the solar emission must be added to the quiet value, and the total flux density becomes
\begin{equation}
F_{\odot}\   =\ F_{\odot q}\ +\ \Delta F\ Z \ \ ,\\
\label{fluxfunc}
\end{equation}
where $Z$ is the sunspot number. The enhancement factor $\Delta F$ is approximated by
\begin{equation}
\Delta F\   =\  1.20\times 10^{-5}\nu^{1.374}\ ,\ \ \ {\rm 30\ -\ 2770\ MHz},\\
\label{fluxfunc1}
\end{equation}
\begin{equation}
\Delta F\   =\ 35.12\ \nu^{-0.5045}\ ,\ \ {\rm 2770\ -\ 10\:000\ MHz},\\
\label{fluxfunc2}
\end{equation}
from data by Smerd (1964) displayed by Kundu (1965). The receiver is close to linear. Deviations from linearity are of the order of a few percent even near the upper limit of the dynamic range (Benz et al., 2005). The measured value $Y(\nu)$ at the frequency channel $\nu$ is thus related to the solar flux $F_{\odot}(\nu)$ by a linear equation
\begin{equation}
Y(\nu)\   \approx\  m(\nu) F_{\odot}(\nu)\ +\ c(\nu)\ \ .
\label{calfunc}
\end{equation}
The calibration constants $m(\nu)$ and $c(\nu)$ are shown in Figure 2 (left and right) for the event presented in Figure 3. The shape of the multiplicative constant $m$ (Figure 2a) defines the observed frequency range of the presented observations. The lower limit around 250 MHz was given by the feed, the upper limit at 900 MHz by a low-pass filter. The oscillations in $m(\nu)$ originate from an impedance mismatch that was later eliminated. Large deviations from the smooth curves were treated as interference and removed similarly to interference in the data described below.

\begin{figure}
\begin{center}
\leavevmode
\mbox{\hspace{0.0cm}\epsfxsize=6.0cm
\epsffile{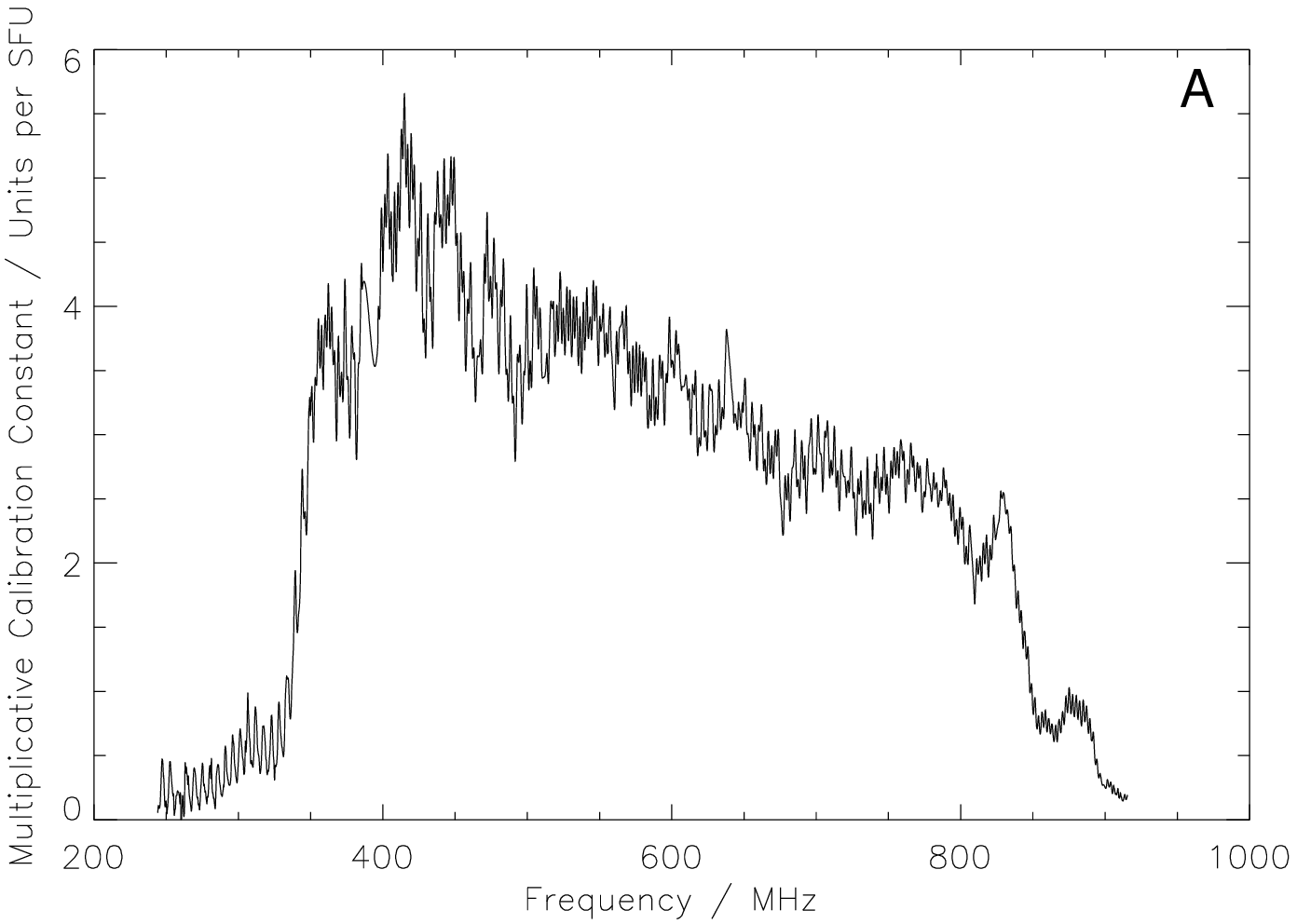}}
\mbox{\hspace{0.0cm}\epsfxsize=6.0cm
\epsffile{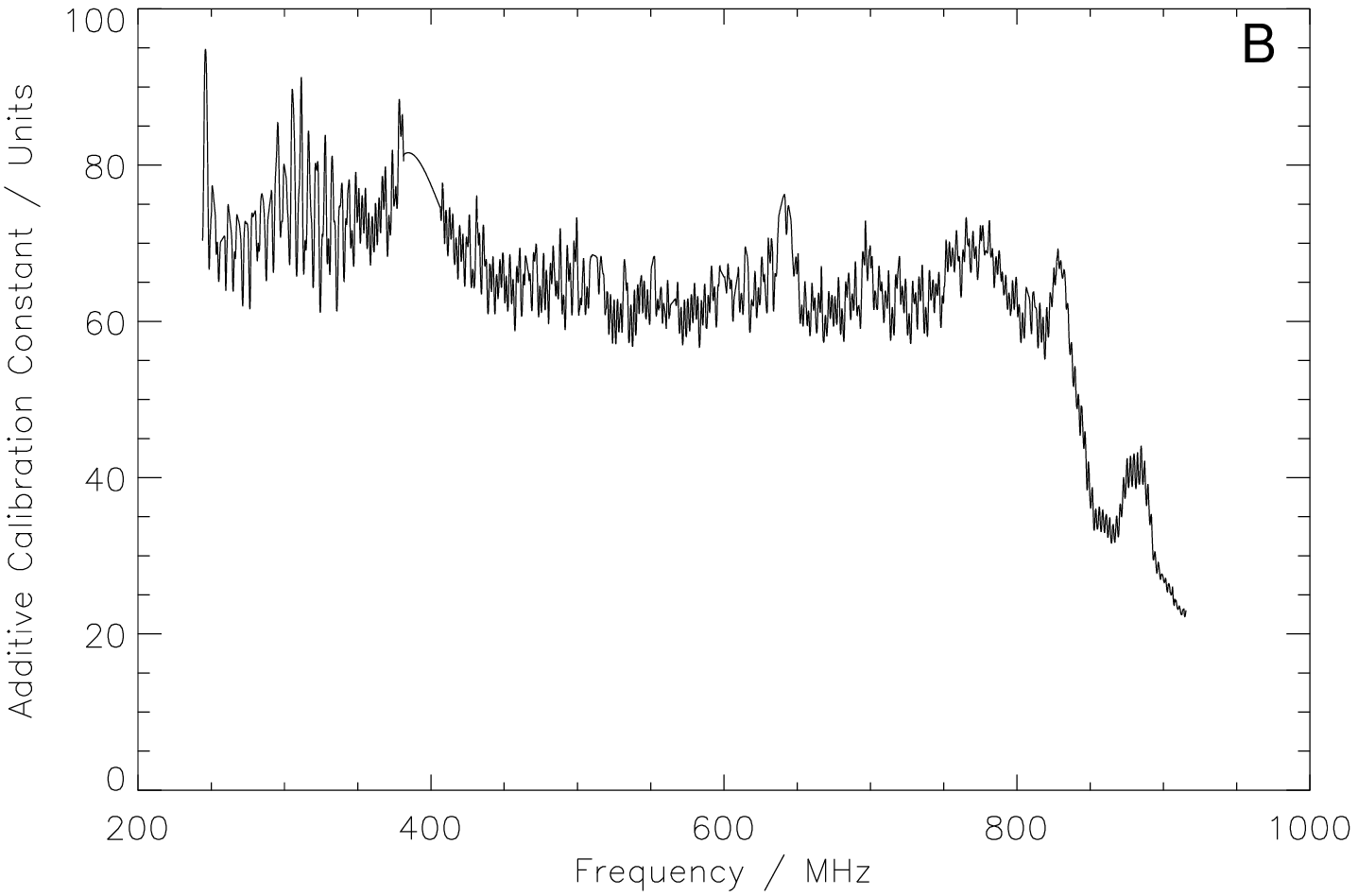}}
\end{center}
\caption[]{{\it A:} Calibration constant $m$ as defined in Equation (\ref{calfunc}) vs. frequency.  {\it B:} Calibration constant $c$  vs. frequency.}  
\label{calib}
\end{figure}

Polarization measurements have been tested with the well defined polarization of down link transmitters on Earth-orbiting satellites. We have also compared our calibration of solar events with observations at the Trieste Solar Radio Observatory (P. Zlobec, private communication). The agreement is satisfactory (better than 20\%). 

The data are calibrated at the Institute of Astronomy at ETH Zurich, after retrieval from the Bleien Observatory. The flux density is logarithmically compressed, circular polarization converted to percent, and the results are permanently stored. Flux density and circular polarization are publicly available.\footnote {http://www.astro.phys.ethz.ch/cgi-bin/showdir?dir=Ph3Observations/bleien/ phoenix3}

\section{First Results}
\begin{figure}
\begin{center}
\leavevmode
\mbox{\hspace{-1.0cm}\epsfxsize=14cm
\epsffile{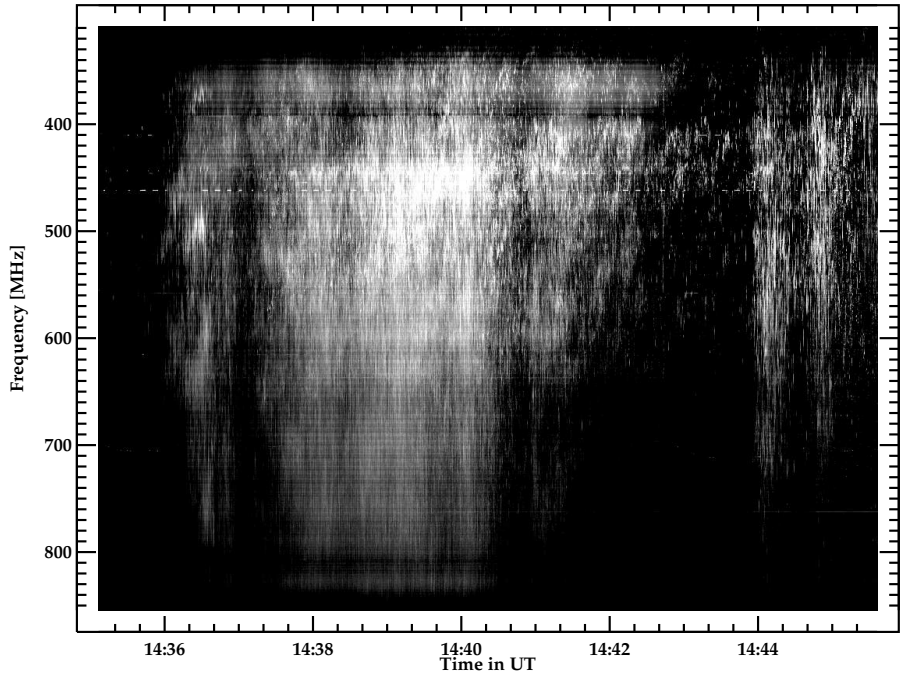}}
\end{center}
\caption[]{Spectrogram of the Phoenix-3 data recorded on 23 August 2005 showing an event that consists of narrowband spikes. The emission before the event is considered as background and subtracted. The result is logarithmically compressed. Horizontal lines (bright or dark) are caused by terrestrial interference. The horizontal ripple results from a temporary impedance mismatch.}  
  \label{spectrogram}
\end{figure}

In summer 2005, Phoenix-3 was tested in the range 0.1 - 1 GHz for solar observations. All channels were recorded without spectral compression, providing a resolution of 61.0 kHz at a time resolution of 100 ms.  On 23 August 2005 an M3 limb flare occurred accompanied by Type IV radio emissions at meter and decimeter wavelengths lasting from 1422.6 to 1652.5 UT. The flare was well observed also by Phoenix-2 and RHESSI (Battaglia and Benz, 2009). Narrowband spikes are observed between 350 - 850 MHz from 14:35:30 to 14:45:00 UT. They are not fully resolved by Phoenix-3 in time, but very well in frequency. The spectrogram shown in Figure \ref{spectrogram} gives an overview.

In Figure \ref{spectrogram} some terrestrial emissions and spurious instrumental features are visible that interfere with the analysis of the spectral characteristics. The instrumental artifacts include variations in sensitivity (that can largely be calibrated out) and spurious peaks at the clock frequency of the FFT device and its harmonics. The latter are stable and affect only one channel, but are nosier than undisturbed channels and must be removed. They are treated the same way as terrestrial interference in two steps: ({\it i}) detection and ({\it ii}) removal.

 As Phoenix-3 undersamples some solar emission in time (see example in Figure \ref{spike}), it is difficult to distinguish interference from solar bursts along the time axis. However, the high spectral resolution allows detecting interference in the frequency axis. The following effects cause spectral variations ({\it i}) An increase from one channel to the next may be due to noise. The combined one $\sigma$ rms fluctuations of the system and the background is of the order of one percent. ({\it ii}) Modulations caused by impedance mismatch are similar or lower. ({\it iii}) Solar bursts may show an increase up to 20\% from one channel
to the next (see below). Thus any spectral increase by more than 50\% can be safely considered to be caused by interference.

\begin{figure}
\begin{center}
\leavevmode
\mbox{\hspace{0.0cm}\epsfxsize=12cm
\epsffile{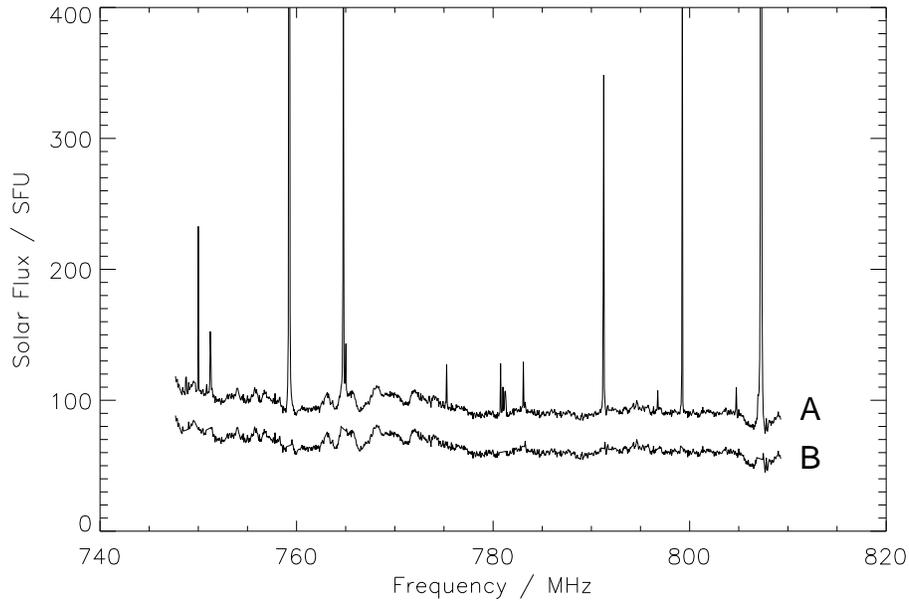}}
\end{center}
\caption[]{{\it A:} Part of an original spectral scan observed with Phoenix-3.  {\it B:}  Same spectrum cleaned by the method described in the text, vertically shifted in flux density for clarity.}  
  \label{fig:4}
\end{figure}

The study of interference revealed that it spreads sometimes beyond one channel and may have several peaks. First the peak with the largest amplitude was identified and the edges of its range located by finding the nearest minima on either side below half power. The peak was removed from the spectra by interpolating the 10 lowest channels of the 100 neighboring channels of the interfered one with a polynomial of fifth order. A tolerance value is defined for each channel from its temporal history of variability. Depending on its value, a channel was removed if the jump to the previous one surpassed between 4 $\sigma$ and 50\%. The process was repeated in each spectrum until the maximum flux was below 3 $\sigma$ above the background. Figure 4 shows an example of the spectrum before and after cleaning. More details on the cleaning algorithm are given by Beverland (2007).
 
\begin{figure}
\begin{center}
\leavevmode
\mbox{\hspace{0.0cm}\epsfxsize=12cm
\epsffile{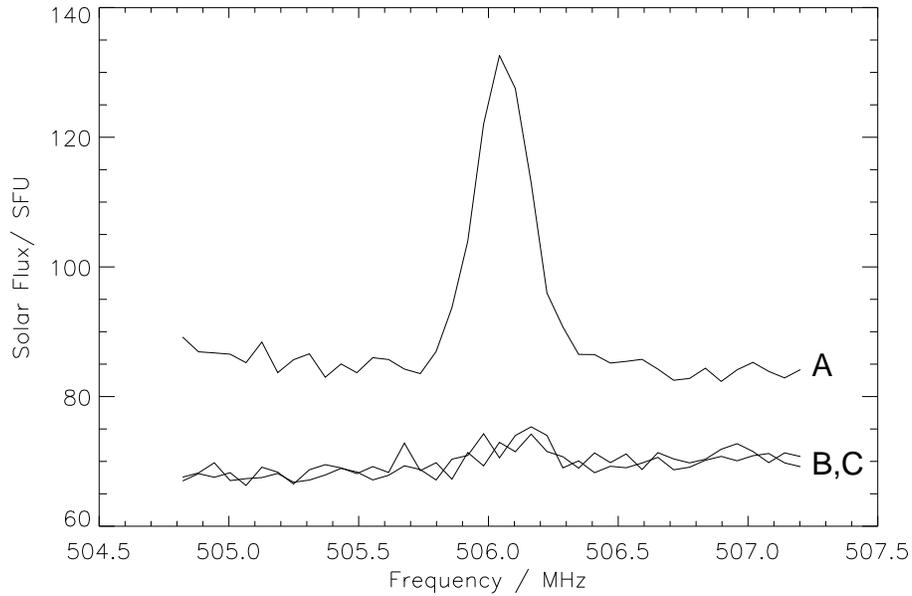}}
\end{center}
\caption[]{{\it A:} Part of a spectrum recorded with Phoenix-3 comprising a single narrowband spike burst. {\it B and C:} The spectra a time step (0.1 s) before and a time step after the spike burst are shown below, vertically shifted for clarity. }  
  \label{spike}
\end{figure}
 
\subsection{Individual spikes}

Figure \ref{spike} shows the close up of an isolated spike and the adjacent spectra. The spectral profile of the spike is resolved by 9 measurement points with a resolution of 61.0 kHz. The FWHM bandwidth is 0.22 MHz amounting to 0.043\% of the central frequency. The spike appears to be fully resolved.

\begin{figure}
\begin{center}
\leavevmode
\mbox{\hspace{-0.30cm}\epsfxsize=8.7cm
\epsffile{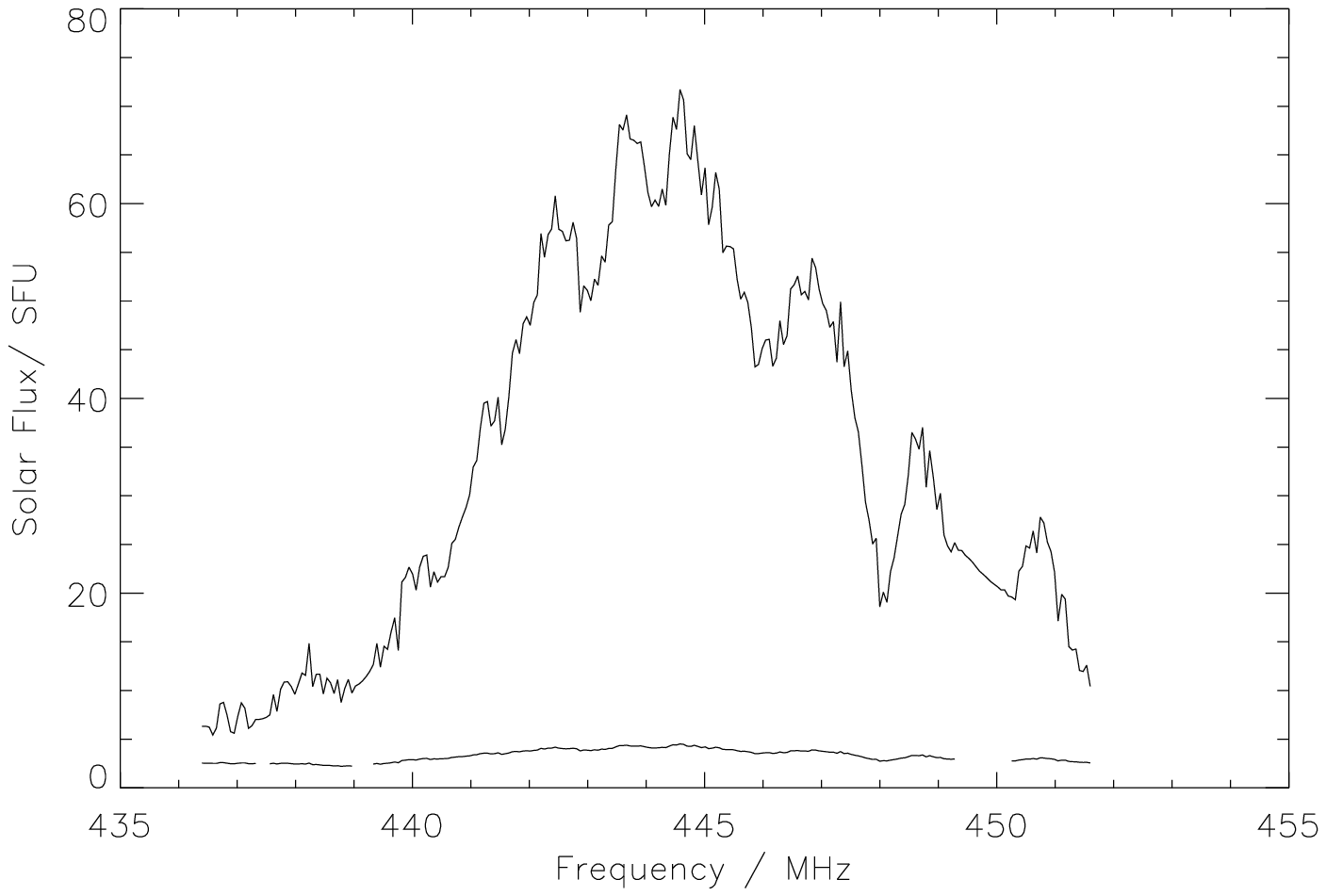}}
\mbox{\hspace{0.0cm}\epsfxsize=4cm
\epsffile{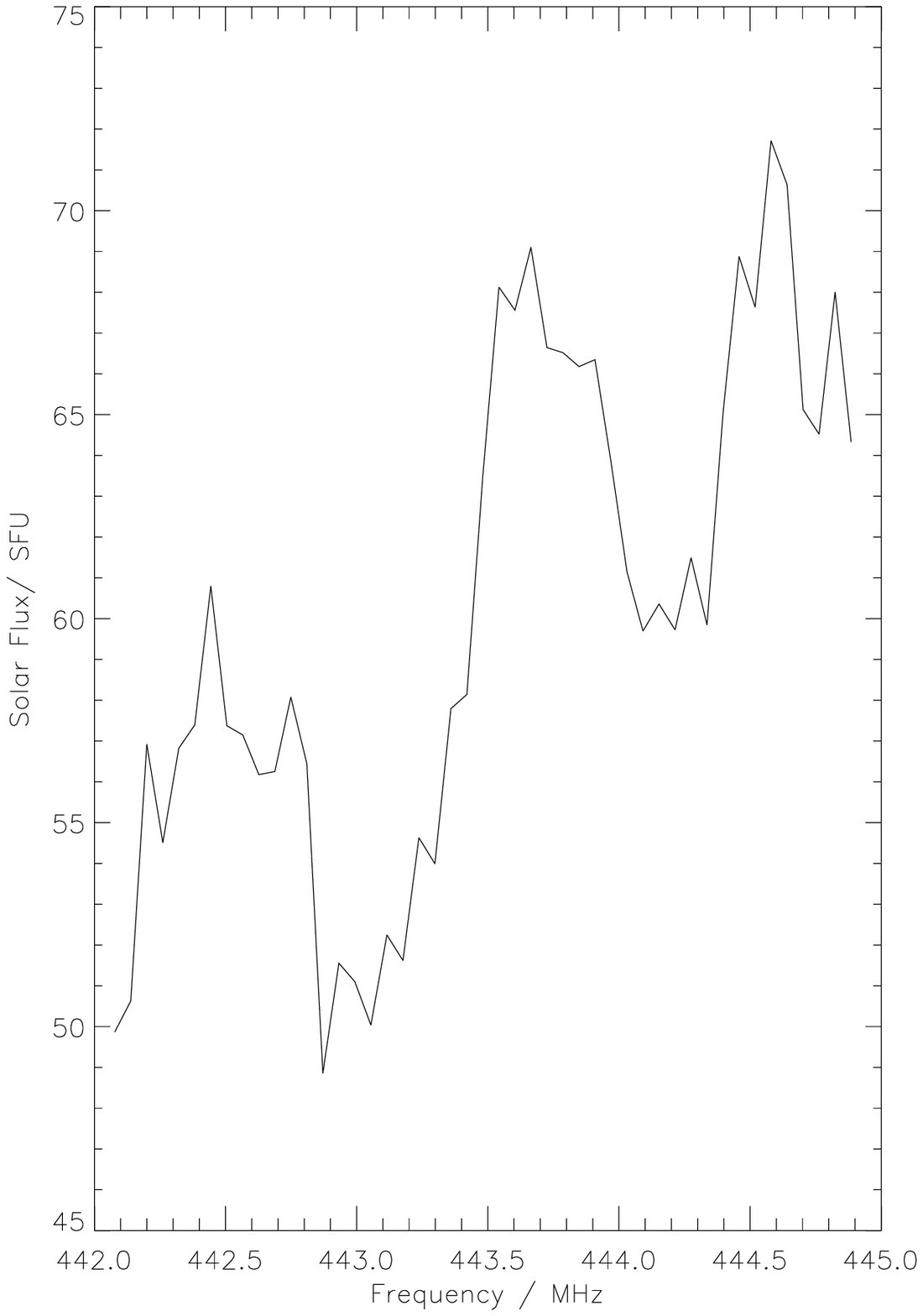}}
\end{center}
\caption[]{Example of steep spectral gradients in the spike emission in a crowded spectrum. {\it Left:} The quiet time noise (5$\sigma$) is shown at the bottom, leaving out the channels with interference. These channels are interpolated in the curve above. {\it Right:} Close-up showing steepest gradient in frequency.}  
  \label{gradient}
\end{figure}

Some other spectral scans are crowded by many spikes and possibly not fully resolved. An example is shown in Figure \ref{gradient}, where the background was subtracted. The rms noise determined before and after the event is shown in Figure 6 left. Its 5 $\sigma$ level varies between 2 and 4 sfu. The flatness of this noise level in frequency indicates that the channels with extremely steep spectral gradients are not interfered before and after the event. The largest jump in Figure 6b amounts to 5 sfu from one channel to the next. It corresponds to a spectral gradient of 80 sfu MHz$^{-1}$ and to an increase by 9\%. The shortest half-power rise is 0.12 MHz. These structures are the narrowest features yet reported in solar radio observations. 

\subsection{Statistical analysis and results}

Individual spikes (see for example Figure \ref{spike}) have a bell shape that can be reduced to a few parameters, such as total flux, bandwidth, and skewness. These characteristics are determined by fitting a modified Gaussian model $F_m(\nu)$ to the observed flux density $F(\nu)$,
\begin{equation}
F_m(\nu)\   =\ F_0  \exp\left( - z^2\over 2\right)[M z + 1] \ +\ {F_0 [A z^2 + B z + C]\over {1 + (z/\alpha)^8}},\\
\label{modGauss}
\end{equation}
where
\begin{equation}
z\ =\ {\nu - \nu_0\over\sigma}\ \ .\\
\label{def_z}
\end{equation}

The first term on the right is a Gaussian modified to allow for skewness. The second term on the right, a modified polynomial, allows for non-Gaussian variations. Its denominator is within 2\% of unity for $z \lapprox 0.5\alpha$  and becomes very large for $z > 2\alpha$. Taking $\alpha = 0.5$ and requiring $\mid M\mid,\ \mid A\mid$, $\mid B\mid$, and $\mid C\mid$ below 0.8, ensures that the second term may contribute to the region surrounding $\nu_0$, but not farther in frequency than about $\sigma$. The optimal values of the seven constants are then found for each spike minimizing $\chi^2$.

Each spectrum was first cleaned of interference. The background was defined by the average of the lowest 5\% of flux values for each channel before and after the event and was subtracted. Spikes are so numerous in the data sample that they often overlap one-another. The solution to this problem is to first identify the various spikes and minimize $\chi^2$ afterwards for the sum of the modeling functions for each spike.

Spikes were identified sequentially - by searching the spectrum for the largest peak first, recording its height, location and approximate width, then removing it and repeating the procedure until no peaks above 20 sfu remained. The recorded values of height, location and width were used as initial conditions for $F_0,\ \nu_0$ and $\sigma$ respectively. The other four constants were initially set to zero prior to optimisation. The optimum seven constants for each spike were found by minimizing $\chi^2$ of the sum of all models in a given spectrum.

The modified Gaussian failed to model some spectral scans, such as shown in Figure \ref{gradient},
revealing internal structure (multiple peaks, steep gradients). The frequency range was restricted to 600 - 850 MHz to reduce effects of interference and crowding. Finally, the fitting resulted in parameters useful for statistical analysis for 1478 spikes.

\begin{figure}
\begin{center}
\leavevmode
\mbox{\hspace{-0.30cm}\epsfxsize=12cm
\epsffile{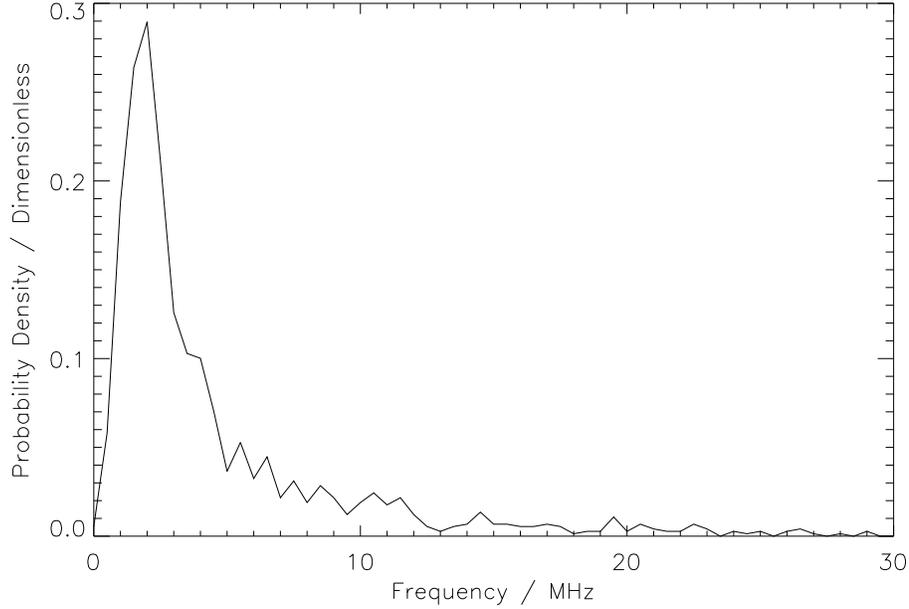}}
\end{center}
\caption[]{Distribution of the FWHM bandwidth of narrowband spikes, normalized to unity. }  
  \label{distr_bandwidth}
\end{figure}

Figure \ref{distr_bandwidth} displays the probability distribution of the spike bandwidth, normalized to unity. The distribution $f(\Delta \nu_{\scriptscriptstyle\rm FWHM})$ follows a power law
\begin{equation}
f(\Delta \nu_{\scriptscriptstyle\rm FWHM})\ =\ 0.93(\pm 0.01) \Delta\nu_{\scriptscriptstyle\rm FWHM}^{-1.7(\pm 0.1)}\ \ \ \ \ [{\rm MHZ}^{-1}].\\
\label{bandwidth}
\end{equation}
The distribution peaks at $\Delta\nu_{\scriptscriptstyle\rm FWHM} = 2$ MHz and drops off below. A bandwidth of 2 MHz corresponds to 32 channels. Therefore, the peak and the drop-off appear to be real. Nevertheless, there are a few spikes having a bandwidth of 0.3 MHz or less.

\begin{figure}
\begin{center}
\leavevmode
\mbox{\hspace{-0.30cm}\epsfxsize=12cm
\epsffile{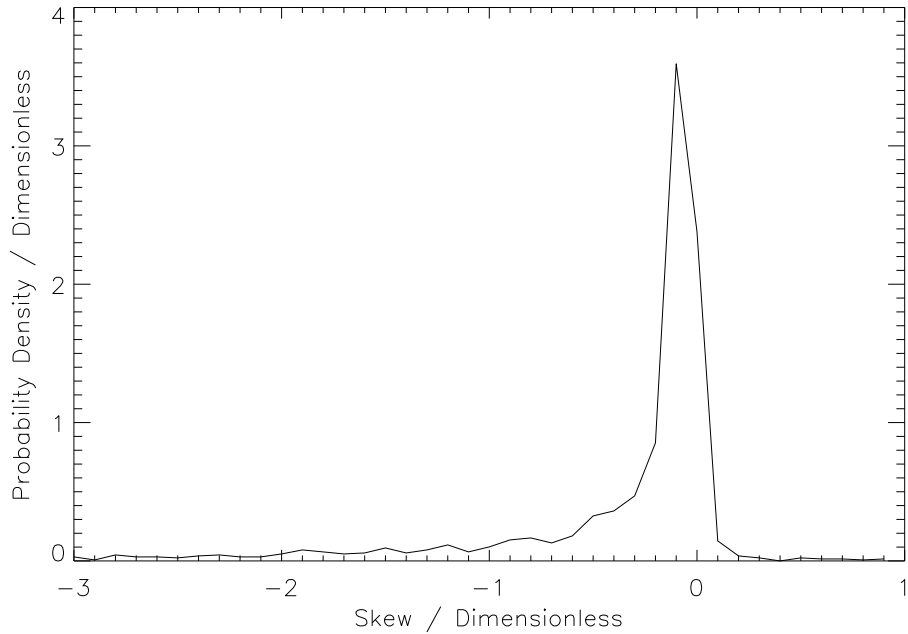}}
\end{center}
\caption[]{Distribution of measured skewness of narrowband spikes. Negative values indicate an enhanced tail to lower frequency.}  
  \label{distr_skew}
\end{figure}

The skewness S is defined by 
\begin{equation}
S\   =\  {{{1\over E} \int_{-\infty}^\infty (\nu - \bar{\nu})^3 F(\nu) d\nu}\over \left( {1\over E} \int_{-\infty}^\infty (\nu - \bar{\nu})^2 F(\nu) d\nu \right)^{3/2}} \ \ ,
\label{skew}
\end{equation}
where 
\begin{equation}
E\   =\  \int_{-\infty}^\infty F(\nu) d\nu \ \ , \\
\label{def_E}
\end{equation}
\begin{equation}
\bar{\nu}\   =\  {1\over E} \int_{-\infty}^\infty \nu F(\nu) d\nu \ \ , \\
\label{def_nu_bar}
\end{equation}
The measured skewness of spikes is not zero. The distribution shown in Figure \ref{distr_skew} peaks at -0.08$\pm 0.01$ and extends to large negative values. The majority of spikes is asymmetric with negative skewness. Negative skew indicates an enhanced tail in frequencies below $\nu_0$.

\begin{figure}
\begin{center}
\leavevmode
\mbox{\hspace{-0.30cm}\epsfxsize=12cm
\epsffile{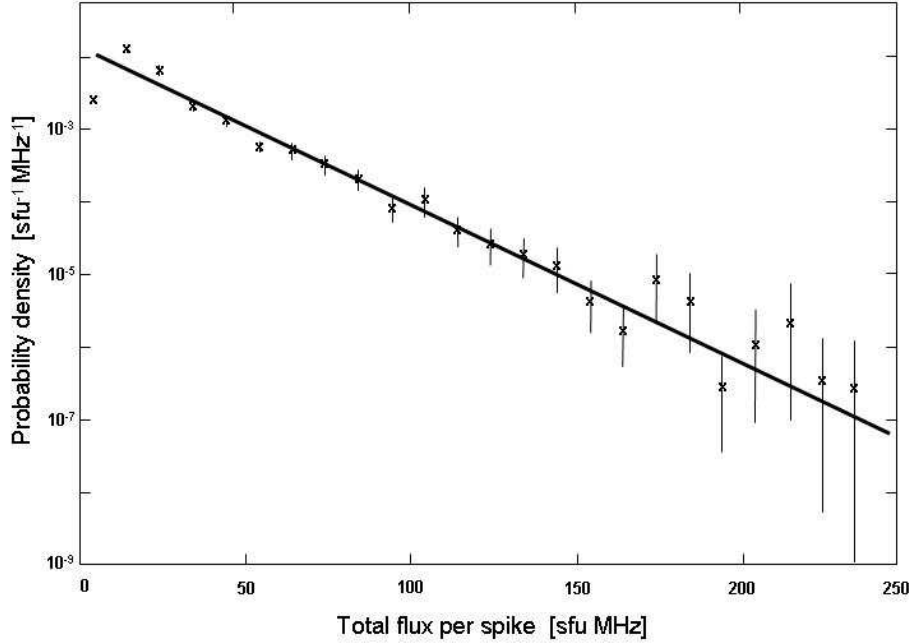}}
\end{center}
\caption[]{Distribution of total flux (flux density integrated in frequency) of narrowband spikes. The line is an exponential fit which is given in Equation (\ref{expofit}). }  
  \label{distr_flux}
\end{figure}
The distribution of spikes in total flux $E$ (Equation \ref{def_E}) is shown in Figure \ref{distr_flux}. It is normalized to probability density. The exponential law,
\begin{equation}
f(E)\   =\  f_0 \exp(-{E\over E_0}) \ \ ,
\label{expofit}
\end{equation}
fits well with the constants $f_0 = 5.0(\pm 0.9)\times 10^{-2}$ sfu$^{-1}$ MHz$^{-1}$ and $E_0 = 20(\pm 2)$ sfu MHz. The drop-off below 15 sfu MHz is a sensitivity effect.

\section{Discussion}
The first observations of Phoenix-3 lead to surprisingly different results from previous reports. We discuss three such findings possibly related to the higher spectral resolution.

\subsection{Bandwidth}
The distribution of bandwidths (Figure \ref{distr_bandwidth}) largely confirms the previous impression that narrowband spikes have no intrinsic bandwidth. The drop-off at 2 MHz, however, indicates that there is a minimum bandwidth, possibly related to the natural width of the emission process (see Section 1). The natural width is usually exceeded by another process broadening spikes to the observed bandwidth. A plausible reason in the case of EMC radiation is given in Equation (2), suggesting that inhomogeneity in the magnetic field determines the bandwidth. For beam instabilities driving Langmuir waves, the corresponding inhomogeneity would be in electron density. 

The narrowest bandwidth in the statistically analyzed band from 600 - 850 MHz is 0.03 \% of the central frequency. It is more than 5 times smaller than previous reports at these wavelengths presented in Section 1. The discrepancy results very likely from the superior spectral resolution of Phoenix-3. 

To achieve the natural width of the emission process, the source must be very homogeneous or very small. For the spike presented in Figure \ref{spike}, Equation (2) requires
\begin{equation}
{\Delta L\over H} \lapprox {\Delta \nu_{\scriptscriptstyle\rm FWHM} \over \nu}  \approx 4.3 \cdot 10^{-4}\ \ \ .
\end{equation}
Assuming that the relevant scale length in density or magnetic field strength, $H$, is $10^8$ cm, the size of the radio source is $\Delta L \approx 10^4$ cm.

\subsection{Skewness}
The systematically negative skewness (Figure \ref{distr_skew}) is a surprising result. The effect is too small to be observed with 1 MHz resolution. It is not evident from single bursts, such as presented in Figure \ref{spike}, without the analysis described in Equation (13). 

Features in solar flare decimeter radiation with pronounced skewness have been reported before. They have been named "tadpoles" after their shape in spectrum, showing an emission "eye" at high frequency, an absorption "body" at frequencies below and an emission "tail" extending to lower frequencies (Slottje, 1972; Magdalenic et al., 2006). However, tadpole structures extend over 5 - 40 MHz, an order of magnitude more than indicated in Figure \ref{distr_skew}. The discovered skewness may represent a general shape at a much lower level. Tadpole structures are extremely rare, and may be the extreme cases of this more general characteristic. This needs to be confirmed in future studies.

\subsection{Total power distribution}
The total flux distribution of spikes (Figure \ref{distr_flux}) is well fit by an exponential over a range exceeding an order of magnitude. Obviously, higher spectral resolution has not resulted in a power-law distribution. The distribution found here is in conflict mainly with the result published by Nita, Fleishman, and Gary (2008) and partially with Meszarosova et al. (2000). The discrepancy may be attributed to the following reasons:
\begin{itemize}
\item Nita, Fleishman, and Gary (2008) studied the distribution of peak flux density, contrary to the total flux distribution here. If duration diminishes with peak flux and the dependence is not a power law, the distribution of total flux may be flattened at low peaks, possibly turning an approximate power law into an exponential.

\item Nita, Fleishman, and Gary (2008) have made a special effort to detect small spikes. 
\item The data presented here are resolved in frequency, contrary to Nita, Fleishman, and Gary (2008), who based their statistics on peaks resolved in time. It is not clear, however, how this would change the distribution and needs further investigation.
\item The event analyzed here is rich in spikes. Such events tend to have exponential distributions as pointed out by Meszarosova et al. (2000).

\end{itemize}
The distribution found here is clearly not caused by limited resolution, but has to be considered as an intrinsic property of the emission process. Nevertheless, Robinson, Li, and Cairns (2004) argue that a log-normal distribution can be well approximated by an exponential law over a limited range of fluxes. They also show that the stochastic growth theory of spikes can be generalized to include quasi-exponentially distributed elementary bursts. The exponential distribution found here is therefore not in contradiction to stochastic wave growth. 

\section{Conclusions}
The use of a multichannel spectrometer, such as Phoenix-3, greatly increases the sensitivity. Compared to its frequency-agile predecessor, Phoenix-2 with 1 MHz resolution, the sensitivity increases by a factor of 3.5 at a 16 times higher spectral resolution. 

The high spectral resolution has additionally facilitated removing interference, which often has a bandwidth of less than 1 MHz. Even for large bandwidth interference, the high resolution helps to reduce the unusable frequency band or allows seeing in-between two interfered bands. On the other hand, the initial digitalization by 8 bit is found insufficient to resolve the strongest interference between 900 and 1000 MHz. Therefore, this spectral range had to be suppressed by a low-pass filter. In the mean time since the first observations in 2005, the frequency range below 1 GHz has seen a number of additional interfering transmitters at the Bleien Observatory. Thus Phoenix-3 had to be reduced to the frequencies above 1 GHz.

Even at full resolution there are still variations from channel to channel larger than the noise (Figures \ref{spike} and \ref{gradient}). However, the bandwidth distribution (Figure \ref{distr_bandwidth}) indicates that Phoenix-3 generally resolves narrowband spikes at its full spectral resolution, but not in routine mode.  Thus future observations at even higher spectral resolution are predicted to not yield steeper gradients and smaller bandwidths. The observed steepest spectral gradients agree with the natural width suggested by theory (see Section 1).

The skewness of spikes towards an enhanced low-frequency tail and their total flux distribution following an exponential law are observational findings requiring theoretical interpretations. The first results from the Phoenix-3 spectrometer demonstrate that the study of solar radio bursts is still up for surprises.

\begin{acknowledgements}
We thank the many people who have also contributed to the successful completion of the FFT implementation on AC240, in particular Dino Zardet (FHNW Windisch), Karl U. Schenk (FHNW Windisch) and Viktor Hunger\-b\"uhler (Acqiris SA, Geneva). Frieder Aebersold has made the mechanical adaptations, and Andr\'e Csillaghy (FHNW Windisch) has contributed software to turn the device into a solar instrument. We thank Paolo Zlobec (Trieste) for help in calibration and Iver Cairns (U. Sydney) for a helpful discussion on flux distributions. The construction of the Phoenix-3 spectrometer is financially supported by the Swiss National Science Foundation (grants 20-113556 and 200020-121676).
\end{acknowledgements}

{} 

\end{article}
\end{document}